\documentclass[12pt,dvips]{article}
\usepackage{moriond}
\usepackage{epsfig}
\usepackage{bm}
\usepackage{color}
\usepackage{graphicx}
\usepackage{amsmath}

\def\Journal#1#2#3#4{{#1} {\bf #2}, #3 (#4)}

\def\EPJA{{\em Eur. Phys. J.} A}

\def\NPA{{\em Nucl. Phys.} A}

\def\PLB{{\em Phys. Lett.}  B}
\def\PRL{\em Phys. Rev. Lett.}
\def\PRC{{\em Phys. Rev.} C}
\def\PRD{{\em Phys. Rev.} D}

\def\JHEP{\em JHEP}
\def\hepph#1{ArXiv:hep-ph/{#1}}

\def\be{\begin{equation}}
\def\ee{\end{equation}}
\def\bea{\begin{eqnarray}}
\def\eea{\end{eqnarray}}

\newcommand{\ud}     {\mathrm{d}}

\newcommand{\Mev}    {\:\mathrm{MeV}}
\newcommand{\gevsq}  {\:\mathrm{GeV}^2}

\newcommand{\ceps}{\varepsilon}

\newcommand{\average}[1]{\left\langle{#1}\right\rangle}
\newcommand{\eq}[1]{Eq.(\ref{#1})}

\newcommand{\sqw}{\sin^2{\theta_W}}

\newcommand{\as}{\alpha_{\scriptscriptstyle S}}

\newcommand{\PMF}{\mathcal{P}_{\mathrm{MF}}}
\newcommand{\Pcor}{\mathcal{P}_{\mathrm{cor}}}

\begin{document}
\vspace*{4cm}
\title{
On the Paschos--Wolfenstein Relationship for Nuclei
 }
\author{S. A. Kulagin%
}
\address{Institute for Nuclear Research, 117312 Moscow, Russia}
\maketitle\abstracts{
Nuclear effects and QCD perturbative corrections to the
Paschos--Wolfenstein relationship are discussed. We argue that perturbative
corrections largely cancel out in this relationship for total cross
sections while the neutron excess correction in heavy nuclei is enhanced
by Fermi motion and nuclear binding effects. These observations are
discussed in the context of NuTeV measurement of the Weinberg mixing angle.
}

The scattering of (anti)neutrino from matter is mediated
by charged $W^+$ or $W^-$ boson (charged current,
CC), or by neutral $Z$ boson (neutral current, NC).
A relation between neutrino--antineutrino asymmetries in the NC and
CC deep-inelastic (DIS) cross sections was derived long ago by Paschos and
Wolfenstein \cite{PW73}
\begin{eqnarray}
\label{pw}
R^- &=&
\frac{
\sigma^\nu_{\mathrm{NC}}-\sigma^{\bar\nu}_{\mathrm{NC}}
}
{
\sigma^\nu_{\mathrm{CC}}-\sigma^{\bar\nu}_{\mathrm{CC}}
}
=\frac12 -\sqw,
\end{eqnarray}
where $\theta_W$ is the Weinberg mixing angle.
The derivation of the Paschos--Wolfenstein relationship (PW) is solely
based on the isospin symmetry and neglects contributions from heavy
quarks. For this reason this relation is exact for an isoscalar target in
a world without heavy quarks. In particular, this means that various
strong interaction effects, including nuclear effects, should cancel out
in $R^-$ for an isoscalar target thus making \eq{pw} a very
good tool for the measurement of the mixing angle in neutrino scattering.

However, in the real world the PW relation is subject to a number of
corrections. In particular, it must be corrected for the effects due to
possible $s-\bar s$ and $c-\bar c$ asymmetries in the target (see e.g.
Refs.~\cite{DFGRS02,mm03}). Furthermore, the targets used in neutrino
experiments are usually heavy nuclei, such as iron in NuTeV
experiment~\cite{nutev-prl}. Heavy nuclei typically have an excess of
neutrons over protons and are non-isoscalar. For a non-isoscalar target
\eq{pw} is not exact and receives various strong interaction corrections
via admixture of the isovector components to $R^-$.
In this paper we address QCD perturbative corrections and nuclear effects
in the PW relationship for non-isoscalar nuclei.

We will discuss (anti)neutrino DIS in the leading twist (LT) QCD
approximation. In this approximation the NC and CC structure functions are
given in terms of parton distribution functions (PDFs). In order to
simplify discussion of isospin effects, we consider the isoscalar,
$q_0(x)=u(x)+d(x)$, and the isovector, $q_1(x)=u(x)-d(x)$, quark
distributions (for simplicity, we suppress the explicit notation for the
$Q^2$ dependence of parton distributions). The calculation of the NC and
CC cross sections, and the PW ratio in the leading order in the strong
coupling constant (LO) is straightforward. The next-to-leading order (NLO)
correction to the PW relation is given in Ref.\cite{DFGRS02} and the
next-to-next-to-leading order (NNLO) correction was calculated in
Ref.\cite{mm03}. The result can be written as
\footnote{\noindent
The total cross sections involve the integration of
the structure functions over the full phase space of $x$ and $Q^2$.
Therefore $\as$ and the moments $x_i^-$ of the parton distributions are
taken at some average scale $Q^2$, which has to be chosen according to
specific experimental conditions. The NNLO coefficient in \eq{pw:cor} is
given in $\overline{MS}$ scheme. We also remark that there is an error in
the $\as$ NLO coefficient in Refs.\cite{DFGRS02,ku03}. I am grateful to K.
McFarland for pointing this out.}
\begin{eqnarray}
\label{pw:cor}
R^- &=& \frac12 - s_W^2 +\delta R^-, \nonumber\\
\delta R^- &=& \left[1-\frac73 s_W^2 + \left(\frac89\frac{\as}{\pi}
+5.34\frac{\as^2}{\pi^2}\right) \left(\frac12-s_W^2\right)\right]
        \left(\frac{x_1^-}{x_0^-}\right),
\end{eqnarray}
where $s_W^2=\sqw$, $\as$ is the strong coupling, and $x_a^- =\int \ud
        x\,x(q_a-\bar q_a)$, with $q_a$ and $\bar q_a$ the distribution
functions of quarks and antiquarks of type $a$. The subscripts 0 and 1
refer to the isoscalar $q_0$ and isovector $q_1$ quark distributions,
respectively. In the derivation of \eq{pw:cor} we expanded in
$x_{1}^-/x_0^-$ and retained only linear corrections. We also neglected
contributions due to possible $s-\bar s$ asymmetry discussion of which can
be found in Refs.\cite{DFGRS02,mm03,ku03}.

Equation (\ref{pw:cor}) applies to any, not necessarily isoscalar, nuclear
target. We observe that $\delta R^-$ is determined by the valence part of
the isovector quark distribution in the target. Complex nuclei, such as
iron, have unequal number of neutrons ($N$) and protons ($Z$) and the
isovector quark distribution is finite in such nuclei. In order to
understand this effect, we first consider a simple approximation which is
often used in processing of DIS data. In particular, we neglect nuclear
effects and view the neutrino scattering off a nucleus as incoherent
scattering off bound protons and neutrons at rest. We denote $q_{a/T}$ as
the distribution of quarks of type $a$ in a target $T$. Then in considered
approximation the nuclear distribution $q_{a{/}A}$ is the sum of the
corresponding quark distributions for bound protons and neutrons
\begin{eqnarray}
\label{nuke:NZ}
q_{a{/}A} &=& Z q_{a{/}p}+ N q_{a{/}n}.
\end{eqnarray}
We apply \eq{nuke:NZ} to the isovector and isoscalar distributions
assuming the isospin invariance of PDFs in the proton and neutron, i.e.
$q_{0/p}(x)=q_{0/n}(x)$ and $q_{1/p}(x)=-q_{1/n}(x)$. Then we have
$q_{0{/}A}(x)=A\,q_{0{/}p}(x)$ and $q_{1{/}A}(x)=(Z-N)q_{1{/}p}(x)$. For
the ratio of average light cone momenta in the isovector and isoscalar
states, which determine $\delta R^-$ in \eq{pw:cor}, we obtain
\begin{eqnarray}
(x_{1}/x_{0})_A=-\delta N (x_{1}/x_{0})_p,
\label{x1:x0}
\end{eqnarray}
where $\delta N=(N-Z)/A$ is fractional excess of neutrons.

It follows from Eq.(\ref{x1:x0}) that the neutron excess correction to
$R^-$ is negative for neutron-rich targets. In order to estimate the
magnitude of this correction for iron target we first neglect $\as$ terms
in \eq{pw:cor} and compute $(x_1^-/x_0^-)_p=0.45$ using the parton
distributions of Ref.\cite{a02} at $Q^2=20\gevsq$. Keeping in mind
application to NuTeV measurement~\cite{nutev-prl} we use $\delta N=0.0574$
reported by NuTeV~\cite{nutev-prd}. Then we have $\delta R^-=-0.013$. This
is a large value on the scale of experimental errors of NuTeV measurement
since $|\delta R^-| \simeq 10\sigma$ (for a discussion of the neutron
excess correction in this context see Ref.~\cite{ku03}). This gives us a
motivation to study various corrections to $R^-$ in more detail.

We first discuss perturbative corrections in \eq{pw:cor} and consider the
difference between NNLO and LO approximations
\begin{equation}
\Delta R^- = \delta R^-(\text{NNLO})-\delta R^-(\text{LO}).
\label{Delta:R}
\end{equation}
If we simply use the LO PDFs and $\as$ of Ref.~\cite{a02} at
$Q^2=20\gevsq$ we obtain from \eq{pw:cor}  $\Delta R^-=-0.0008$, which is
about 6\% of the LO value of $\delta R^-$. However this calculation is not
fully consistent, since PDFs as well as the value of $\as$ in \eq{Delta:R}
should correspond to the order of perturbative calculation. It is possible
to take into account this effect using the results of analysis of
Ref.~\cite{a02}, which provides PDFs to different order up to the NNLO
approximation. If we do so we observe that the terms in the right side of
\eq{Delta:R} almost cancel each other leading to $\Delta R^-\simeq
0.7\cdot 10^{-4}$. This value is the order of magnitude less than the
result of a naive calculation (note also that the sign of the correction
has changed). The reason for this cancellation is that $\as$ terms in
\eq{pw:cor} turned out to be balanced by perturbative effects in PDFs
which cause the decrease in the ratio $x_1/x_0$ for valence quarks in the
proton from 0.457 (LO) to 0.434 (NNLO).

In order to verify that this cancellation is not accidential we performed
similar analysis for $Q^2=10$ and $100\gevsq$. We have respectively
$\Delta R^-=-1.7\cdot 10^{-5}$ and $1.9\cdot 10^{-4}$. These values
indicate that the cancellation seems to be systematic. It must be also
noted that such a small value of $\Delta R^-$ suggests that the magnitude
of perturbative correction to $R^-$ is within the variations of $\delta
R^-$ due to PDF uncertainties of Ref.~\cite{a02}. Summarizing, we conclude
that the LO calculation provides a good approximation of $R^-$.

Now we turn to the discussion of nuclear effects in $R^-$.
In order to improve on \eq{nuke:NZ}, we consider nuclear binding and Fermi
motion effects (for which we will use the abbreviation FMB) in terms of the
convolution model of nuclear parton distributions (see, e.g.,
Refs.~\cite{akv85,ku89,kpw94}). Then \eq{nuke:NZ} should be replaced by
\begin{eqnarray}
q_{a/A} = \average{q_{a/p}}_p + \average{q_{a/n}}_n,
\label{nuke:qA}
\end{eqnarray}
where the  two terms in the right side are
the quark distributions in bound protons and neutrons averaged
with the proton and neutron nuclear spectral functions, respectively.
Similar equation can also be written for
antiquark distribution.
The explicit expression for the averaging in \eq{nuke:qA} is
(see \cite{ku89,kpw94})
\begin{eqnarray}\label{nuke:av}
x\langle q_{a/p}\rangle_p &=& \int \ud\ceps\ud^3\bm{k}\,
        \mathcal{P}_p(\ceps,\bm{k})
        \left(1+\frac{k_z}{M}\right)
        x'q_{a/p}(x'),\\
x'&=&\frac{Q^2}{2k\cdot q}=\frac{x}{1+(\ceps+k_z)/M}.
\label{xprim}
\end{eqnarray}
The integration in \eq{nuke:av} is taken over the energy  and momentum of
bound protons (we separate the nucleon mass $M$ from the nucleon energy
$k_0=M+\ceps$). The quantity $\mathcal{P}_p(\ceps,\bm{k})$ is the nuclear
spectral function which describes the distribution of bound protons over
the energy and momentum. In \eq{nuke:av}, the $z$-axis is chosen in the
direction opposite to the momentum transfer $q=(q_0,0_\perp,-|\bm{q}|)$,
and $x'$ is the Bjorken variable of the bound proton with four-momentum
$k$. Equation similar to \eq{nuke:av} also holds for neutrons with the
obvious replacement of the spectral function and quark distributions. The
spectral functions $\mathcal{P}_{p}$ and $\mathcal{P}_{n}$ are normalized
to the proton and neutron number, respectively.
%
%

For the isoscalar and isovector nuclear parton distributions we obtain
from Eq.(\ref{nuke:qA})
\begin{subequations}
\label{qA:01}
\begin{eqnarray}
\label{nuke:q0}
q_{0/A} &=& \average{q_{0/p}}_0,\\
q_{1/A} &=& \average{q_{1/p}}_1,
\label{nuke:q1}
\end{eqnarray}
\end{subequations}
where the averaging is respectively performed with isoscalar and isovector
spectral functions, $\mathcal{P}_0=\mathcal{P}_{p}+\mathcal{P}_{n}$ and
$\mathcal{P}_1=\mathcal{P}_{p}-\mathcal{P}_{n}$.

The isoscalar and isovector spectral functions $\mathcal{P}_0$ and
$\mathcal{P}_1$ are very different in complex nuclei.
In an isoscalar nucleus with equal number of protons and
neutrons one generally assumes vanishing $\mathcal{P}_1$
\footnote{\noindent
It must be commented that this statement is violated by a number of
effects even in the $Z=N$ nuclei. The finite difference
between the proton and neutron spectral functions is generated by
Coulomb interaction and isospin-dependent effects in the
nucleon--nucleon interaction. The discussion of these effects goes
beyond the scope of this paper and we leave this topic for future studies.
}
and nuclear effects are dominated by the isoscalar spectral function.
In a nuclear mean-field model, in which a nucleus is viewed as Fermi gas
of nucleons bound to self-consistent mean field, the spectral function can
be calculated as
\begin{equation}
\label{spfn:MF}
\PMF(\ceps,\bm{p})=\sum_{\lambda<\lambda_F}
n_\lambda \left|\phi_\lambda(\bm{p})\right|^2
\delta(\ceps - \ceps_\lambda),
\end{equation}
where $\phi_\lambda(\bm{p})$ are the  wave functions of the
single-particle level $\lambda$ in nuclear mean field and $n_\lambda$ is
the number of nucleons on this level. The sum in \eq{spfn:MF} runs
over occupied single-particle levels with energies below the Fermi level
$\lambda_F$.
Equation (\ref{spfn:MF}) gives a good approximation to nuclear spectral
function in the vicinity of the Fermi level, where the excitation energies
of the residual nucleus are small.
As separation energy $|\ceps|$ becomes higher, \eq{spfn:MF} becomes
less accurate.
High-energy and high-momentum
component of nuclear spectrum can not be described by the mean-field model
and driven by correlation effects in nuclear ground state as witnessed by
numerous studies. We denote this contributions to the spectral function as
$\Pcor(\ceps,\bm{p})$.

In a generic nucleus the spectral function $\mathcal{P}_1$ determines the
isovector nucleon distribution. We now argue that the strength
of $\mathcal{P}_1$ is peaked about the Fermi surface. It is reasonable to
assume that $\Pcor$ is mainly isoscalar and neglect its contribution to
$\mathcal{P}_1$. Then $\mathcal{P}_1$
is determined by the difference of the proton and neutron
mean-field spectral functions. If we further neglect small differences
between the energy levels of protons and neutrons then $\mathcal{P}_1$
will be determined by the difference in the level occupation numbers
$n_\lambda$ for protons and neutrons. Because of Pauli principle, an
additional particle can join a Fermi system only on an unoccupied level.
In a complex nucleus all but the Fermi level are usually occupied (the
Fermi level has a large degeneracy factor). Therefore,
$\mathcal{P}_1$ is determined by the contribution from the Fermi level and
we can write
\begin{equation}
\label{spfn_1}
\mathcal{P}_1=(Z-N)|\phi_{F}(\bm{p})|^2\delta(\ceps-\ceps_F),
\end{equation}
where $\ceps_F$ and $\phi_F$ are the energy and the wave function of
the Fermi level.
In a nucleus with a large number of particles one can use
the Fermi gas model to evaluate the wave function $\phi_F$. In this model
$|\psi_F(p)|^2 \propto \delta(p_F-p)$, where $p_F$ is the Fermi momentum,
and we have
\begin{eqnarray}
\mathcal{P}_1 &=& (Z-N)\delta(p-p_F)\delta(\ceps-\ceps_F)/(4\pi p_F^2).
\label{DeltaP:FG}
\end{eqnarray}

We now apply these equations to calculate the binding and momentum
distribution effects on average quark light-cone momenta in the isoscalar
and isovector quark distributions. Integrating Eq.(\ref{nuke:av}) over $x$
and keeping the terms to order $\ceps/M$ and $\bm{k}^2/M^2$ we have
\begin{subequations}\label{x01:av}
\begin{eqnarray}
\frac{x_{0/A}}{A} &=& \left(1+\frac{\ceps_0+\frac23 T_0}{M}\right)x_{0/p},\\
\frac{x_{1/A}}{A} &=& -\delta N
                       \left(1+\frac{\ceps_F+\frac23 T_F}{M}\right)x_{1/p},
\end{eqnarray}
\end{subequations}
where $\ceps_0$ and $T_0$ are the separation and kinetic energy per
nucleon averaged with the isoscalar nuclear spectral function
$\mathcal{P}_0$ and $T_F=p_F^2/(2M)$. In order to quantitatively estimate
this effect we observe that the energy of the Fermi level $\ceps_F$ equals
the minimum nucleon separation energy. For the iron nucleus we take
$\ceps_F= -10\Mev$ and $p_F=260\Mev$ (the corresponding energy $T_F=
36\Mev$). In order to calculate the isoscalar parameters $\ceps_0$ and
$T_0$ we use the model spectral function of Ref.\cite{CS96} which takes
into account both the mean-field and correlated contributions (see also
~\cite{KS00}). We find that the naive neutron excess correction by
\eq{x1:x0} should be increased by the factor 1.055.

We now discuss these results in the context of NuTeV
effect~\cite{nutev-prl}. We assume that the weak mixing angle can be
calculated from \eq{pw:cor} in terms of experimental $R^-$.
In particular,
we are interested in the variation of $s_W^2$ because of nuclear effects
and effects of higher order in $\as$, since NuTeV analysis was carried out
in LO approximation without nuclear effects. The correction $\Delta s_W^2$
is apparently given by the difference between the corrected and
uncorrected experessions for $\delta R^-$
\begin{equation}
\Delta s_W^2 = \delta R^-(\text{NNLO+FMB})-\delta R^-(\text{LO}).
\label{Delta:sw}
\end{equation}
Perturbative corrections are largely canceled out in the difference as
discussed above and the resulting value $\Delta s_W^2=-0.00065$ is
practically saturated by nuclear effects. With certain care this value can
be viewed as a correction to the value of the Weinberg angle $\sqw$
measured by NuTeV \cite{nutev-prl}.%
\footnote{
It should be remarked in this context that total
cross sections were not measured in NuTeV experiment. For
this reason a more detailed analysis of differential cross sections with
the proper treatment of experimental cuts and (anti)neutrino flux is
needed. }

In summary, we discussed perturbative QCD corrections to the PW
relationship together with nuclear binding and Fermi motion effects. A
cancellation of QCD perturbative corrections to the PW relationship for the
total cross sections has been observed. We found a negative correction due
to nuclear effects to the PW relationship for the total cross sections for
neutron-rich targets and estimated this effect on the Weinberg angle of
NuTeV measurement.

\section*{Acknowledgments}
This work was supported in part by the RFBR project no. 03-02-17177.
I am grateful to the organizers of the Moriond EW2004 meeting for support
and warm hospitality.


\end{document}